\documentclass[twocolumn,preprintnumbers,amsmath,amssymb,prl]{revtex4}
\usepackage{graphicx}% Include figure files
\usepackage{dcolumn}% Align table columns on decimal point
\usepackage{bm}% bold math

\begin{document}
\title{Tunable bandgap and magnetic ordering 
by adsorption of molecules on graphene}
\author{Julia Berashevich and Tapash Chakraborty}
\email{chakrabt@cc.umanitoba.ca}
\affiliation{Department of Physics and Astronomy, 
University of Manitoba, Winnipeg, Canada, R3T 2N2}

\begin{abstract}
We have studied the electronic and magnetic properties of graphene
and their modification due to the adsorption of water and other
gas molecules. Water and gas molecules adsorbed on nanoscale 
graphene were found to play the role of defects which facilitate 
the tunability of the bandgap and allow us to control 
the magnetic ordering of localized states at the edges. 
The adsorbed molecules push the wavefunctions corresponding 
to $\alpha$-spin (up) and $\beta$-spin (down) states of 
graphene to the opposite (zigzag) edges. 
This breaks the sublattice and molecular point
group symmetry that results in opening of a large bandgap. 
The efficiency of the wavefunction displacement depends 
strongly on the type of molecules adsorbed. 
\end{abstract}
\maketitle

Monolayer graphene \cite{nov1} with its high electron mobility, 
unique magnetic phenomena \cite{expt2}, and unusual 
relativistic-like properties \cite{tapash} has 
generated an upsurge of activities in materials science. 
Materials exhibiting magnetic properties are in great demand 
for applications in nanoscale electronics and spintronics. 
Most magnetic materials are metals, where ferromagnetism (which 
results from an imbalance between the spin-up and spin-down 
unpaired electrons) is often destroyed by thermal fluctuations. 
In carbon systems the magnetic properties are stable even at 
room temperatures. The nonmetalicity of carbon systems makes 
them biocompatible and therefore are ideal for a wide range of 
applications not only in nanoelectronics and spintronics 
\cite{cohen,rudb,cho}, but also in medicine and biology. 
In most carbon systems the origin of magnetism is unclear, but 
in graphene it is expected to be the result of spin ordered 
states localized at the edges \cite{cohen1,kim,nak,lee,pisani,fer,fuj}.

The absence of a bandgap is however a major hindrance for graphene's 
application in nanoelectronics. Various mechanisms for opening a gap 
have been developed that involve breaking of certain symmetries 
in graphene by defects \cite{pisani}, an applied bias \cite{cohen,rudb}, 
and interaction with other materials \cite{zhou}. 
Noting that the conductivity of the carbon systems 
\cite{geim,wehling} is extremely sensitive to adsorption of gas 
molecules due to the charge exchange between them, we assume that 
a more controllable way to induce interactions that break a symmetry 
of graphene would be gas adsorption. The charge exchange 
between the adsorbed molecules and graphite was found to be 
rather low \cite{ort}, and donor or acceptor behavior is exhibited 
depending on the type of molecules adsorbed. However, since gas 
adsorption changes the electronic properties of graphene due to 
the charge exchange between them, we suspect that adsorption should 
affect the localization of electronic states along the edges and can
facilitate opening of a gap. 

%Figure  1
\begin{figure}
\includegraphics[scale=0.30]{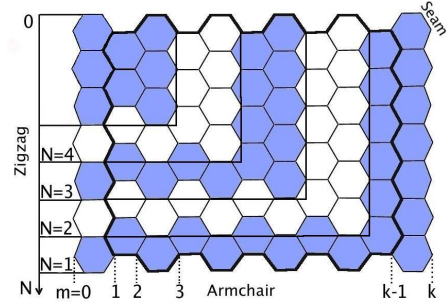}
\caption{\label{fig:fig1} (color online). Nanoscale graphene structures, where 
$N$ is the number of rings along the zigzag edges and 
$m$ is the along the armchair edges.}
\end{figure}

Here we study the influence of adsorption of water on the electronic 
and magnetic properties of graphene by molecular mechanical methods using 
the spin-polarized density functional theory with semilocal gradient 
corrected functional (UB3LYP/6-31G \cite{b3lyp}) in the Jaguar 6.5 program \cite{jaguar}. 
The van der Waals interactions, 
which impact the interaction between graphene and 
adsorpant, are not considered in DFT, that lead to underestimation 
of the adsorption energy \cite{ort}. However, the DFT is reliable 
for investigation of the influence of the adsorption 
on the electronic properties of the graphene. 
We chose the nanoscale graphene structures as shown in Fig.~1. 
The number of rings along the armchair edges was varied as $m=0\ldots 
k$ or $m=1 \ldots (k-1)$ to take into account different geometries 
of the graphene {\it seams} (indicated in Fig.~1), where $k$ was 
always odd. The number of carbon rings along the zigzag edges $N$ was 
varied from 3 to 7.

For a nanoscale graphene the highest possible symmetry is D2h, i.e., 
the planar symmetry with an inversion center, where all four seams are 
characterized by the same boundary conditions and exhibit the lowest 
energy in the system due to the confinement effect. The optimization 
process makes these seams structurally identical with the same density 
distribution. Therefore, the seams have the same spin order and play 
the role of the center for spin ordering along the edges (see 
Fig.~2(a)). The resulting spin polarization exhibits a ferromagnetic 
ordering between the seams and between the zigzag edges, but a mixed 
ordering along the zigzag edges. Most molecular orbitals including 
the highest occupied and the lowest unoccupied (HOMO and LUMO) exhibit 
zero spin density, because they have the same density distribution 
for the $\alpha$- and $\beta$-spin states, but with opposite 
sign. For the HOMO and LUMO each spin state is equally delocalized 
over the two zigzag edges. The molecular orbitals contributing to a 
non-zero spin density are located far from HOMO-LUMO gap and they 
are not localized along zigzag edges. The spin density and spin 
polarization for graphene of size $N$=4 are presented in Fig.~2 (a).
For nanoscale graphene optimized without any symmetry, with  
the planar symmetry C1 or with the C2v symmetry, the seams are not constrained to be identical. 
That makes the zigzag edges exhibit the lowest energy and as a result,
a ferromagnetic ordering of the spins along the zigzag edges with
a bandgap of $\sim$ 1.6 eV for C2v symmetry and C1 symmetry,
instead of $\sim$ 0.5 eV for D2h symmetry. 
In these cases, the HOMO and LUMO are 
characterized by the $\alpha$- and $\beta$-spin states localized on 
the opposite zigzag edges. Although the state of graphene with D2h symmetry 
is comparable in energy to the states of
C2v and C1 symmetries, the difference between them is already 
less then 0.02 eV for the $N\ge$6 and $m\ge7$. This implements the 
high metastability of the ground state of C2v symmetry.
Therefore, D2h is chosen here in order to study the influence 
of adsorption of gas and water molecules on both spin ordering and the bandgap.

%Figure 2
\begin{figure*}
\includegraphics[scale=0.57]{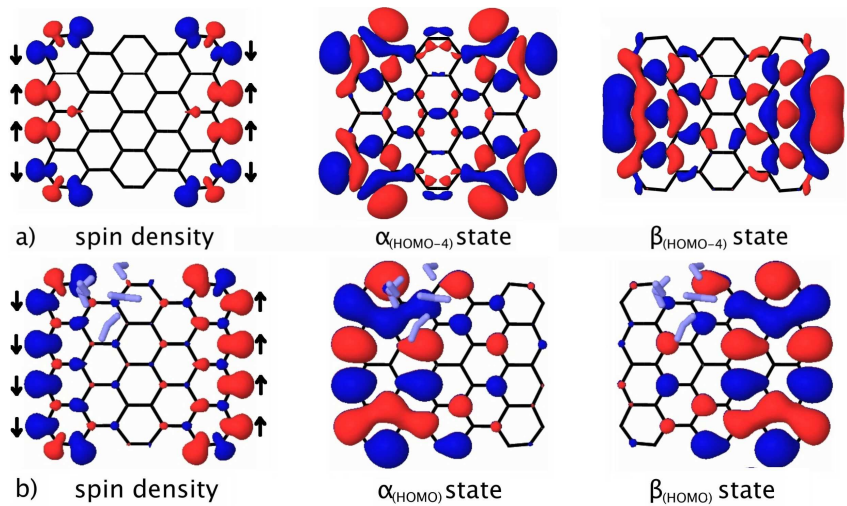}
\protect\caption{\label{fig:fig2} (color online). Spin polarizations in nanoscale 
graphene of size $N$=4 and $m=0 \ldots k$: The spin density and 
electron density for the $\alpha$-spin (up) and $\beta$-spin (down)
states are plotted for isovalues of $\pm0.01$ e/\AA$^{3}$.
For the spin density, different colors 
indicate the $\alpha$- and $\beta$-spin states, while for the electron 
density distribution ($\alpha$- and $\beta$- states) the different colors 
corresponds to different signs of the molecular orbital lobes.
(a) The electron density distribution in pure graphene
for $\alpha$-spin ($E_{\mathrm{HOMO-4}}=-7.119$ eV) and 
$\beta$-spin ($E_{\mathrm{HOMO-4}}=-7.295$ eV) states 
for the fourth orbital below the highest occupied molecular orbital 
(HOMO-4). In pure graphene, a non-zero spin density is obtained only 
for an even number of carbon rings along the zigzag edge 
($N=4,6, \ldots$). 
(b) The electron density distribution of the $\alpha$-spin state 
($E_{\mathrm{HOMO}}=-5.386$ eV) and the $\beta$-spin state 
($E_{\mathrm{HOMO}}=-5.270$ eV) for the HOMO for graphene with 
adsorbed water. The water cluster plays the role of a defect pushing 
the $\alpha$- and $\beta$-spin wavefunctions to the opposite 
edges.}
\end{figure*}

The D2h symmetric graphene structures with an even number of rings 
along the zigzag edges $N=4,6$ and the number of rings at the armchair 
edges that are varied as $m=0\ldots k$, are found to be the only case 
with a non-zero spin density (Fig.~2 (a)). For these structures and 
sizes, the atomic wave functions from the opposite seams cancel 
each other at the center of the edge, where they are out of phase.
In that case, the $\alpha$- and $\beta$-spin states are spatially 
separated which leads to non-zero spin density for certain 
orbitals. The $\alpha$ state is localized on the seams, while the 
$\beta$ state at the center of the zigzag edges (see Figure 2 (a)). We 
found this effect for several molecular orbitals that are far from 
the HOMO-LUMO gap, while other orbitals including the HOMO and LUMO
have the zero spin density. These states are non-degenerate, that 
diminish with increasing structure size. For example, the energy 
difference between the $\alpha$- and $\beta$-spin states for 
HOMO is 0.0764 eV if $N$=4 and 0.00625 eV if $N$=6. This can explain 
the experimental observation of the single local density peak in 
the valence or the conduction band for large graphene structures 
\cite{exp1,exp2}. For an odd number of carbon rings along the zigzag 
edges $N=3,5,7$ and $m=0\ldots k$, the spin density at the edges 
is close to zero because the $\alpha$- and $\beta$- 
degenerate states exhibit almost the same electron density 
distribution over all molecular orbitals. In a structure 
where the armchair edge is varied as $m=1 \ldots (k-1)$, we 
found the opposite behavior of the spin degeneracy: Degenerate 
$\alpha$ and $\beta$ spin states are observed for the 
even ring number on the zigzag edges, while for 
odd $N$ the states are non-degenerate. 

Our most striking results are for the case of adsorbed water 
molecules on nanoscale graphene. The interaction between 
water and graphene is weak because the attractive forces between 
the water molecules within a cluster are stronger than the attractive 
force of the graphene surface. Therefore, several water molecules 
will tend to form a water cluster, that was studied for graphene structure of sizes 
$m=0 \ldots k$ and $N$=3,4,5,6. A single water molecule on 
the graphene surface is localized at the center of a carbon 
ring, $\sim$\ 3.0 \AA\ above the graphene surface. The optimized 
position of the water molecule on the graphene surface corresponds 
to its dipole moment being perpendicular to the graphene surface 
and directed inward, which provides an effective charge donation 
from the oxygen to the graphene through the unshared pair of $p$ 
electrons of the oxygen. When we increase the number of water 
molecules, this water molecule becomes a ``cluster link" 
to the graphene, i.e., the extra water molecules are connected to 
the cluster link by the hydrogen bonds. Therefore, the charge donation 
to graphene occurs mostly through oxygen in the cluster link, 
while acceptance by the cluster is through the hydrogen 
atoms. For the graphene structure with $N\leq 4$, where the 
confinement effect is strong, the interaction between the armchair 
edge and the water cluster leads to unrolling of the water cluster 
in such a way that one termination side is attached to the armchair 
edge and the other -- the cluster link -- to the center 
of a carbon hexagon. The attachment to the 
armchair edge is a result of its low energy in comparison to the 
zigzag edge \cite{koskinen}. The reconstruction of the zigzag edge
can lead to lowering of its energy that would change the 
direction of cluster unrollment. For $N\ge 4$ the water cluster is formed 
at the center of graphene. 

Water can be treated as a defect on the otherwise defect-free 
graphene surface. Figure 2 (b) shows that this defect repels 
the wavefunctions of the molecular orbitals corresponding to 
$\alpha$- and $\beta$-spin states towards opposite zigzag 
edges of the nanoscale graphene. This leads to spin ordering 
of the localized electrons on the edges such that the $\alpha$-spin 
state is localized on one sublattice while the $\beta$-spin state 
on the other sublattice. The non-zero spin density and repulsion of 
the wavefunctions by water were observed for all 
types of graphene structures independent of size. Just as for pure 
graphene, the size is found to be crucial only to preserve the 
non-degeneracy of the $\alpha$- and $\beta$-spin states. The defects 
affect the wavefunction distribution only for the nearest carbon 
rings. This means that, depending on the graphene size and the 
size of the water cluster, the wavefunctions can be repelled 
locally, or over the whole structure. For a graphene size of $N=6$, 
the water molecules build a cluster close to the graphene center. 
Then the $\alpha$- and $\beta$-spin wavefunctions are pushed 
toward the edges only for the localized states along 
the line of the water location (Fig.~3). 

\begin{figure}
\includegraphics[scale=0.37]{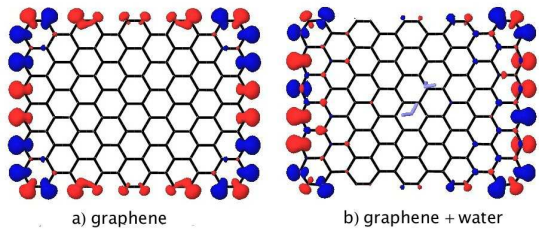}
\caption{\label{fig:fig3} (color online). Spin polarizations in nanoscale 
graphene of size $N$=6 and $m=0 \ldots k$. The spin densities 
are plotted for isovalues of $\pm0.01$ e/\AA$^{3}$:
(a) in pure graphene; (b) for graphene with adsorbed water.}
\end{figure}
 
The repulsion of the $\alpha$- and $\beta$-spin wavefunctions 
to opposite zigzag edges discussed above, breaks the molecular 
and sublattice symmetries and opens a gap. The magnitude of 
the gap directly depends on the interaction of the graphene 
surface with water, particularly on the charge exchange and the 
association energy. For a single water molecule, if the dipole moment 
of the water molecule is directed inward (toward the graphene 
surface), the gap increases by up to $\sim$ 2.0 eV, but if the 
dipole moment is directed outward then the charge donation 
from oxygen to graphene does not occur and the gap opens 
by only up to 0.8 eV. For the inward directed dipole, the 
gap slightly decreases with increasing distance between the 
water molecule and the graphene surface. For the water cluster, 
the cluster link usually is a donor and its interaction 
with the graphene surface defines the association of the water 
cluster to graphene and the efficiency of opening a gap. The 
orientation of the cluster link to the graphene surface 
is changed due to the intrinsic interaction within the water 
cluster. Therefore, the association energy of the water 
cluster on graphene and hence the graphene gap oscillates 
depending on the orientation of the cluster link to the 
graphene surface and correlation between attractive forces inside 
the water cluster and attractive forces of the graphene surface. 
For a cluster containing three water molecules, the association 
energy has a global minimum and the same minimum is observed 
for the graphene gap (Fig.~4 (a)). With an increase of the 
graphene size, the gap  decreases ($\sim$ 1.2 eV for $N\ge$ 6) due to 
the vanishing of the confinement effect and the gap oscillation diminishes 
(the dashed lines in Fig.~4 (a)). The energy gap of 1.2 eV 
is in perfect agreement with earlier investigation of the 
bandgap of nanoribbons with ferromagnetic ordering along the zigzag 
edges by the B3LYP method \cite{rudb}. 
We predict that using a water detergent would 
be a way to reduce the attractive forces inside the water cluster 
and thereby increase the water-graphene interactions. This would 
allow one to tune the energy gap in graphene and improve the 
repulsion of the wavefunctions towards the opposite zigzag edges.

\begin{figure}
\includegraphics[scale=0.44]{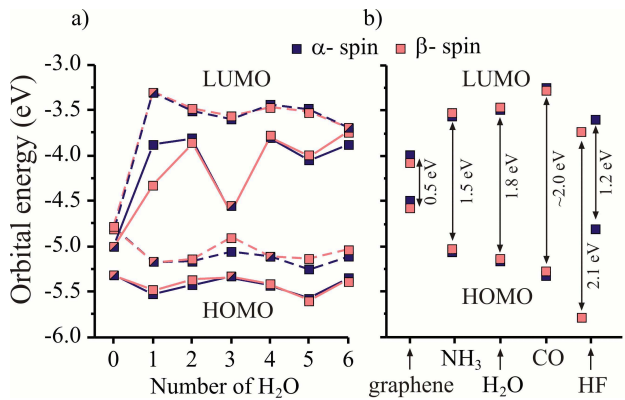}
\caption{\label{fig:fig5} (color online). The effect of adsorption on the gap 
in nanoscale graphene. (a) The HOMO and LUMO energies
for graphene of size $N$=3 (solid line) and $N$=5 (dashed line) 
versus the number of adsorbed water molecules. 
(b) Influence of adsorption of different adsorbed molecules 
on the bandgap of the nanoscale graphene of size $N$=4.}
\end{figure}

Opening of the gap is a result of breaking the sublattice and molecular
symmetries of graphene, but the gap size can vary 
depending on the interaction between the adsorpant and the 
graphene surface. Therefore, we also investigated the possibility 
to tune the energy gap by changing the type of adsorbed 
molecules. The HOMO-LUMO gap for nanoscale graphene with adsorbed 
NH$_3$, H$_2$O, HF, or CO molecules are presented in Fig.~4 (b). 
Since the gap opens due to the symmetry breaking, the interaction 
of nanoscale graphene with any molecule leads to a gap opening for 
up to $\sim$ 1.5-2.0 eV. The ability of the adsorbed molecules to 
push the wavefunctions of $\alpha$- and $\beta$-spin states to edges 
is found to differ between the type of molecules chosen and 
the HF and CO were least effective. Moreover, the HF adsorption leads
to spin-selective semiconductor behavior of the graphene. 

In summary, the adsorbed molecules on graphene are found 
to act as defect centers which alters the magnetism 
through the repulsion of the wavefunctions corresponding to $\alpha$- and 
$\beta$-spin states to the opposite zigzag edges. 
This results in a bandgap opening and stabilization 
of the state with parallel spin ordering along the zigzag edges and 
anti-parallel one between the opposite edges.
The reversible breaking of the graphene symmetry and opening of a gap due to
adsorption, as obsevred in experiments \cite{zhou,dong}, 
would open a new direction in controlling the electronic properties of graphene.
In particular, transition from metallic to semiconductor state by
molecular adsorption is a promising route to
develop fast on-off switching devices \cite{dong}, 
super-sensitive gas sensors \cite{geim,wehling} and FET transistor \cite{lin},
where adsorption can help ambipolar-to-unipolar conversion. 
For FET application, the graphene can operate similar to carbon nanotubes.
The gap opening in graphene induced by adsorption can provide 
ambipolar-unipolar conversion of FET, as a result of 
occurrence of a significant Schottky barrier at metal/graphene contact, 
leading to modification of the conducting channel operation.
The spin-polarized state of graphene stabilized by adsorption 
provides an opportunity to manipulate spin current
by an in-plane electric field applied across the zigzag edges \cite{cohen}.
The applied field increases a bandgap for one-spin state,
while suppressing the other, thereby generating half-metallicity in graphene.
This effect can be used for developing 
spin-related electronic devices \cite{cohen,cho,gun},
such as a spin-valve \cite{cho} and digital memory devices \cite{gun}.

The work was supported by the Canada Research Chairs
Program and the NSERC Discovery Grant.\\

\end{document}